\documentclass[aps,prd
,preprint,tightenlines,
nofootinbib,showpacs]{revtex4}
\usepackage{amssymb,latexsym}
\usepackage{amsmath,amsbsy,bbm}
\usepackage{epsfig,bm}
\usepackage{graphicx,comment}
\DeclareMathOperator{\tr}{tr}

\begin{document}
\def\a{{\alpha}}
\def\b{{\beta}}
\def\d{{\delta}}
\def\D{{\Delta}}
\def\X{{\Xi}}
\def\e{{\varepsilon}}
\def\g{{\gamma}}
\def\G{{\Gamma}}
\def\k{{\kappa}}
\def\l{{\lambda}}
\def\L{{\Lambda}}
\def\m{{\mu}}
\def\n{{\nu}}
\def\o{{\omega}}
\def\O{{\Omega}}
\def\S{{\Sigma}}
\def\s{{\sigma}}
\def\th{{\theta}}

\def\ol#1{{\overline{#1}}}

\def\Aslash{A\hskip-0.45em /}
\def\Dslash{D\hskip-0.65em /}
\def\Dtslash{\tilde{D} \hskip-0.65em /}

\def\CPT{{$\chi$PT$\,$}}
\def\QCPT{{Q$\chi$PT}}
\def\PQCPT{{PQ$\chi$PT}}
\def\tr{\text{tr}}
\def\str{\text{str}}
\def\diag{\text{diag}}
\def\order{{\mathcal O}}

\def\cF{{\mathcal F}}
\def\cS{{\mathcal S}}
\def\cC{{\mathcal C}}
\def\cB{{\mathcal B}}
\def\cT{{\mathcal T}}
\def\cQ{{\mathcal Q}}
\def\cL{{\mathcal L}}
\def\cO{{\mathcal O}}
\def\cA{{\mathcal A}}
\def\cQ{{\mathcal Q}}
\def\cR{{\mathcal R}}
\def\cH{{\mathcal H}}
\def\cW{{\mathcal W}}
\def\cM{{\mathcal M}}
\def\cD{{\mathcal D}}
\def\cN{{\mathcal N}}
\def\cP{{\mathcal P}}
\def\cK{{\mathcal K}}
\def\Qt{{\tilde{Q}}}
\def\Dt{{\tilde{D}}}
\def\St{{\tilde{\Sigma}}}
\def\cBt{{\tilde{\mathcal{B}}}}
\def\cDt{{\tilde{\mathcal{D}}}}
\def\cTt{{\tilde{\mathcal{T}}}}
\def\cMt{{\tilde{\mathcal{M}}}}
\def\At{{\tilde{A}}}
\def\cNt{{\tilde{\mathcal{N}}}}
\def\cOt{{\tilde{\mathcal{O}}}}
\def\cPt{{\tilde{\mathcal{P}}}}
\def\cI{{\mathcal{I}}}
\def\cJ{{\mathcal{J}}}

\def\eqref#1{{(\ref{#1})}}

\preprint{LLNL-JRNL-514075}

\title{Finite isospin density probe for conformality}

\author{Michael I. Buchoff}
\email[]{buchoff1@llnl.gov}
\affiliation{%
Physical Sciences Directorate, 
Lawrence Livermore National Laboratory,
Livermore, California 94550, USA}

\date{\today}


\begin{abstract}
 A new method of employing an isospin chemical potential for QCD-like theories with different number of colors, number of fermion flavors, and in different fermion representations is proposed.  The isospin chemical potential, which can be simulated on the lattice due to its positive definite determinant gives a means to probe both confining theories and IR conformal theories without adjusting the lattice spacing and size.  As the quark mass is reduced, the isospin chemical potential provides an avenue to extract the chiral condensate in confining theories through the resulting pseudoscalar condensate.  For IR conformal theories, the mass anomalous dimension can be extracted in the conformal window through ``finite density" scaling since the isospin chemical potential is coupled to a conserved current.  In both of these approaches, the isospin chemical potential can be continuously varied for each ensemble at comparable costs while maintaining the hierarchy between the lattice size and lattice spacing.   In addition to exploring these methods, finite volume and lattice spacing effects are investigated.

\end{abstract}
\maketitle

\section{Introduction}                                        

The study of strongly coupled beyond the Standard Model (BSM) theories through lattice gauge theory has seen a surge of activity within the last few years as a result of advancements in computational technology and the advent of the LHC.  In order to truly understand the non-perturbative dynamics that would lead to electroweak symmetry breaking in these technicolor theories \cite{Weinberg:1979bn,Susskind:1978ms} (for reviews, see Ref.~\cite{Chivukula:2000mb,Lane:2002wv}), numerical lattice calculations are necessary.  While not completely ruled out, Òscaled upÓ theories of QCD with two flavors and three colors have been disfavored due to a larger than observed S-parameter \cite{Peskin:1990zt,Peskin:1991sw}, and a chiral condensate too small to account for the flavor hierarchy when extended technicolor and flavor changing neutral current bounds are considered \cite{Eichten:1979ah,Dimopoulos:1979es}.  However, the dynamics of strongly coupled gauge theories with different number of flavors, colors, and fermion representations are still largely unknown.  One proposed resolution for these experimental discrepancies is walking technicolor \cite{Holdom:1981rm,Yamawaki:1985zg,Appelquist:1986an,Appelquist:1998xf,Hsu:1998jd,Kurachi:2006mu,Dietrich:2006cm}, which is thought to both yield an enhanced condensate and parity doubling.  For fixed representation and colors, these scenarios are believed to set in for number of flavors just below the onset of the conformal window, after which spontaneous symmetry breaking no longer occurs. The majority of the lattice calculations to date focus on addressing these situations non-perturbatively.

There have been a multitude of lattice studies involving asymptotically free QCD-like theories.   The majority of calculations have explored three-color theories with fermions in the fundamental representation at various number of flavors \cite{Appelquist:2007hu,Deuzeman:2008sc,Deuzeman:2009mh,Fodor:2009wk, Bilgici:2009kh,Appelquist:2009ty,Appelquist:2009ka,Hasenfratz:2009ea, Appelquist:2010xv,Hasenfratz:2010fi, Hayakawa:2010yn,Fodor:2011tu}. However, there have also been multiple studies of two color fundamental \cite{Bursa:2010xn}, two color adjoint \cite{Catterall:2008qk,Catterall:2009sb,DelDebbio:2009fd,DelDebbio:2010hx,DeGrand:2011qd,Giedt:2011kz,Catterall:2011zf}, and three color sextet \cite{DeGrand:2008kx,DeGrand:2009et,DeGrand:2009hu,Kogut:2010cz,Kogut:2011ty}.   While great progress has been made in each of these theories, much debate still remains largely due to the systematic uncertainties that result from discretization choices, non-zero quark masses, and finite volume effects.  Also, unlike QCD, there is little phenomenology to guide these calculations.  Thus, these calculations are heavily reliant on effective field theory (EFT) extrapolations or techniques for critical phenomena to attain information in the chiral limit.  For these techniques to be valid, however, small fermion masses are required.

In this work we explore the effect of introducing a new dimensionful parameter to the action, namely an isospin chemical potential, and how this parameter can help extract the chiral condensate in the chiral limit for confining theories and the mass anomalous dimension in the conformal window.  The isospin chemical potential, which couples to the conserved isospin current, has been studied extensively in QCD \cite{Son:2000xc,Son:2000by} and has been simulated on the lattice \cite{Kogut:2002tm,Kogut:2002zg,Kogut:2004zg,deForcrand:2007uz}.  Isospin chemical potentials have also been explored as a probe for low-energy scattering parameters in meson-baryon systems \cite{Bedaque:2009yh} and two-point functions for mesons \cite{Akemann:2008vp}.  In addition to not having a sign problem, it has several other beneficial properties.  First, for a confining theory with chiral symmetry breaking, an isospin chemical potential leads to a pseudoscalar condensate, provided the chemical potential is larger than its critical value.  The scale of this condensate is set by the confinement scale and can be related to the chiral condensate in certain limits.   Second, unlike the fermion mass term, the isospin chemical potential is coupled to a conserved current, which results in the chemical potential not receiving any renormalization.  This fact, along with a modest value for the mass, allows for one to extract the mass anomalous dimension via ``finite density scaling"  similar to the extraction of critical exponents from finite size scaling.  A similar approach of exploiting the non-renormalization of the chemical potential to extract a critical index was performed in Ref.~\cite{Hands:1992ck}.  Additionally, studies have been performed on both chemical potentials and conformal phase transitions, along with free energy \cite{Lenaghan:2001sd, Mojaza:2010cm}.

The organization of this paper is as follows.  In Sec.~\ref{sec:Iso}, the isospin chemical potential is reviewed for both continuum QCD and lattice QCD.  In Sec~\ref{sec:Confine}, these results are generalized to confining theories and two methods are presented for extracting the chiral condensate in the chiral limit, namely, the pion mass and the pseudoscalar condensate.  In Sec.~\ref{sec:Conform}, universal finite density scaling functions, similar to the finite size scaling functions examined in Ref.~\cite{DeGrand:2009mt,DelDebbio:2010jy,DelDebbio:2010ze} are derived for theories with an IR fixed point and methods for obtaining the mass anomalous dimension via finite density scaling are presented.   In Sec.~\ref{sec:Lat_Art}, an estimate of the lattice artifacts due small quark masses in the presence of an isospin chemical potential is given.

\section{Isospin Chemical Potential} \label{sec:Iso}            
For each global symmetry/conserved quantity that a system possesses, one can add a term to the Lagrangian which consists of a chemical potential coupled to the resulting conserved current.  This term, which corresponds to the energy required to add an additional particle that carries one unit of the associated conserved charge, ultimately allows for exploration of the system at finite density.  Physically, the system of the greatest interest is finite baryon density where a baryon chemical potential is coupled to the current of the conserved baryon number.  However, due to to a complex fermion determinant in Euclidean space, the usual Monte Carlo lattice techniques cannot be applied to complex representation fermions\footnote{Baryon chemical potentials are applicable to a host of theories that are in the real or pseudoreal representation, such as two color theories, adjoint representation calculations, or $SO(N_c)$ gauge theories \cite{Cherman:2010jj,Cherman:2011mh}, due to a positive definite fermion determinant.  While this could be a very interesting approach to many BSM calculations, we do not cover this scenario here.}.  In this work, we will explore a chemical potential coupled to a different conserved current, namely an isospin chemical potential.   As long as the theory possesses an $SU(2)$ subgroup of the flavor symmetry, this approach is applicable.

\subsection{Continuum QCD}                %
The Lagrangian for a two flavor QCD with an isospin chemical potential coupled to the third isospin matrix is given by

\begin{equation}
\label{eq:L_Cont}
\cL = \ol \psi \Bigg[ i\Big(\Dslash + i\mu_I\gamma_0 \frac{\tau^3}{2}\Big) - \cM \Big ] \psi,
\end{equation}
where $\Dslash$ is the $SU(N_c)$ covariant derivative, $\cM = \diag(m_q,m_q)$, $\mu_I$ is the isospin chemical potential, and the fermion fields $\psi$ represent a pair of quarks in an isospin doublet.  When $\mu_I = m_q = 0$, the action has an $SU(2)_L \otimes SU(2)_R$ chiral symmetry which breaks spontaneously to the vector subgroup, $SU(2)_V$.  Turning on the chemical potential term, which behaves like the time component of a uniform gauge field, leads to the action being invariant under $U(1)_L \otimes U(1)_R$ which spontaneously breaks to $U(1)_V$ (turning on the mass term gives an explicit breaking to this symmetry).  This $U(1)_V$ symmetry can be best understood as a circle in the $\tau^1$-$\tau^2$ plane perpendicular to the $\tau_3$ term in the Lagrangian.  

One of the interesting outcomes to having an isospin chemical potential is the possibility of forming a pseudoscalar condensate \cite{Son:2000xc,Son:2000by}.  This condensate comes about when the the chemical potential is larger than a certain critical value and the $U(1)_V$ symmetry is spontaneously broken in the direction of the condensate.  To understand this process, it is best to work with a small quark mass and small chemical potential in the chiral regime (namely, $m_q \sim \mu_I \ll \L_\chi$, where $\Lambda_\chi$ is the chiral symmetry breaking scale).  The leading order (LO) Lagrangian in chiral perturbation theory (\CPT) is given by \cite{Gasser:1983yg,Gasser:1984gg}

\begin{equation}
\label{eq:L_Cont_CPT}
\cL = \frac{F^2}{8}\Big[\tr(D_\mu \S D^\mu \S^\dag) + 2 B \;\tr(M^\dag \S + \S^\dag M) \Big],
\end{equation}
where the normalization $F \sim 130\; \text{MeV}$ and  $M = \diag(m_q,m_q)$.  $\S$ is the matrix of the Nambu-Goldstone fields which can be parameterized as $\S = \exp(i\alpha\;\mathbf{n}\cdot\bm{\tau})= \cos \a + i \mathbf{n}\cdot\bm{\tau}\sin\a$ and $B$ is related to the chiral condensate as the quark mass goes to zero,
\begin{equation}\label{qq_cond }
B = \lim_{m_q\rightarrow 0} \frac{ | \langle \bar{q} q \rangle |}{F^2}. 
\end{equation}
The isospin chemical potential enters in the covariant ``gauge" derivative, $D_\nu \S = \partial_\nu \S + i [\mu_I \frac{\tau^3}{2}\d_{\nu 0}, \S]$.
Expanding Eq.~\eqref{eq:L_Cont_CPT}, one arrives at the potential
\begin{equation}
\label{eq:L_Cont_Pot}
V(\S) = \frac{F^2}{8}\Big[\frac{\mu_I^2}{4}\tr(\big[\tau^3,\S][\tau^3,\S^\dag]\big) -2 B m_q \tr( \S + \S^\dag) \Big].
\end{equation}
\begin{figure}[b]
   \centering
   \includegraphics[width=120mm]{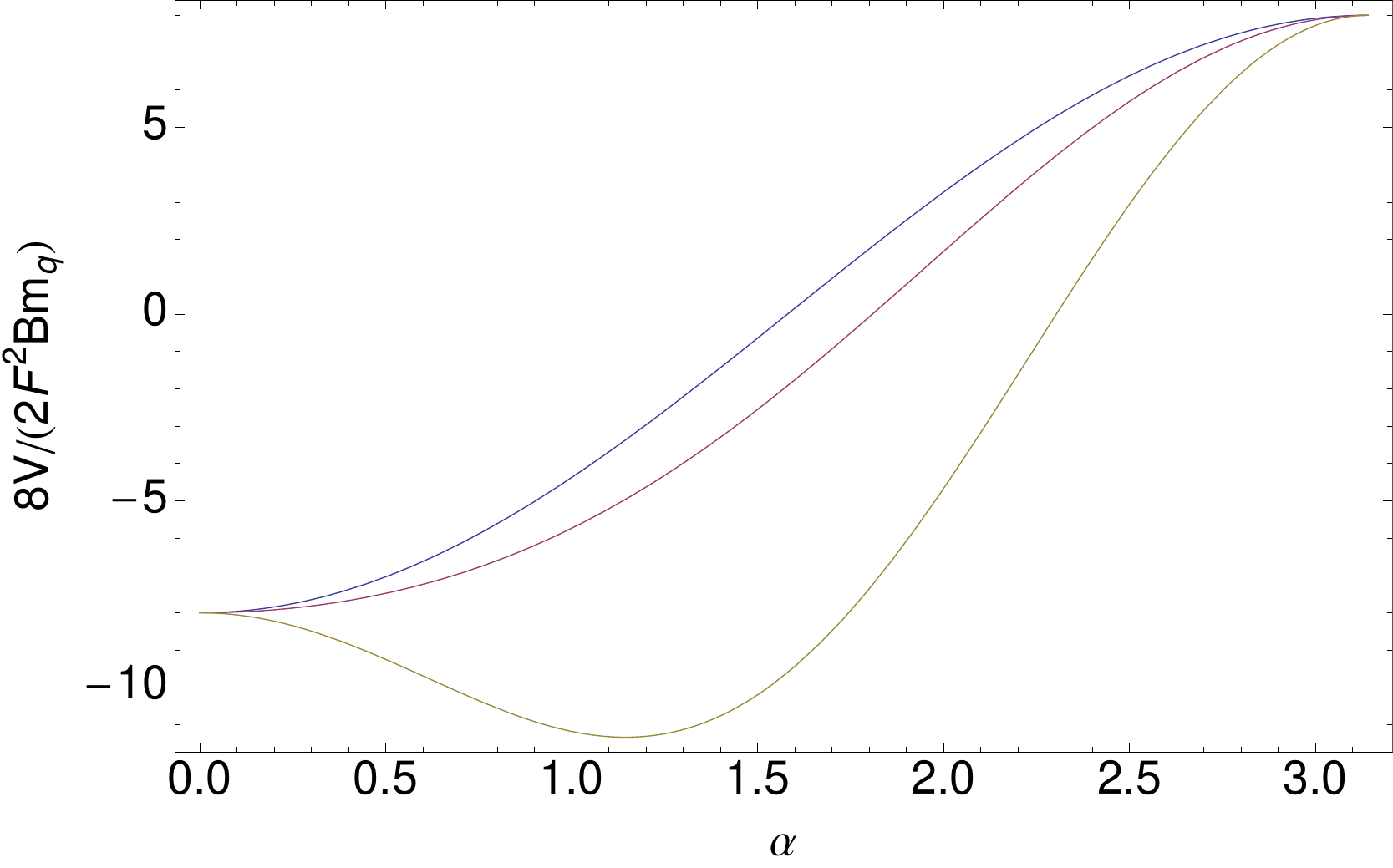} 
   \caption{Chiral potential vs. $\alpha$ for values of $\mu_I$ below (blue), equal (red), and above (yellow) $2Bm_q$.  When above the critical value, the potential develops a minimum at a non-zero value of $\a$, which defines the vacuum expectation value in the condensed phase.}
   \label{fig:Chiral_Pot}
\end{figure}
The potential as a function of $\alpha$ for several isospin chemical potential values is shown in Fig.~\ref{fig:Chiral_Pot}.  When $\mu_I < 2Bm_q$, the potential increases from its value at $\a = 0$ and vacuum remains unchanged. When $\mu_I > 2Bm_q$, the potential forms a new minimum and a condensed phase is favored. In terms of the parameter $\alpha$, the minimum of the potential is given by\footnote{In order to reduce confusion throughout this work, we leave the leading order vacuum pion mass as $2Bm_q$ (Gell-Mann-Oakes-Renner relation).  The quantity $m_\pi$ will be reserved for the pion mass measured in the presence of the pion condensate.}
\begin{equation}
\label{eq:Pot_Min}
\cos \alpha = \frac{2Bm_q}{\mu_I^2}.
\end{equation}
As a result, the vacuum expectation value of the Nambu-Goldstone fields is given by
\begin{equation} \label{eq:pi_cond}
\S_0=
\begin{cases}
1, & | \mu_I | < 2Bm_q, \\
\exp [ i \alpha \, \mathbf{n}\cdot\bm{\tau} ], &
|\mu_I| > 2Bm_q
\end{cases}
,\end{equation}
where $\mathbf{n}$ is the direction of the spontaneously symmetry.  This new alignment will alter the chiral effective field theory and will consequently affect observables.  This is accounted for in the \CPT Lagrangian by expanding $\S$ about this new minimum, $\S = \xi_0 \tilde{\S}\xi_0$, where $\xi_0^2 = \S_0$.  This feature can be used to pick off low-energy coefficients, as proposed in Ref.~\cite{Bedaque:2009yh}, and is explored for confining theories in Sec.~\ref{sec:Confine}.
  
\subsection{Lattice QCD}                %
When extending the continuum formulation of a field theory to a discretized simulation, there are often several subtleties that need to be understood.  The first aspect is how the finite isospin density system maintains a positive definite determinant while the baryon chemical potential does not.  One condition that can ensure this property for a theory with an even number of flavors is that the determinant of the action is real (multiplying an even number of real determinants is greater than or equal to zero).  One relation that ensures a lattice action maintains a real determinant is if the discretized action, $M$, obeys the $\gamma_5$-hermiticity condition
\begin{equation}\label{eq:G5_Herm}
M^\dag = \gamma^5 M \gamma^5 \quad \longrightarrow \quad \det M = \det M^\dag.
\end{equation}
Consequently, the determinant of the product $\det M^\dag M = \det M^\dag \det M = |\det M|^2$ is positive definite.  The inclusion of any chemical potential term of the form $\ol \psi \gamma_0 \psi$ will invalidate this relation.  However, this is not the only condition to lead to a real determinant.  One can include a flavor matrix and arrive at the relation 
\begin{equation}\label{eq:tG5_Herm}
M^\dag = \tau_i\gamma^5 M \tau_i\gamma^5 \quad \longrightarrow \quad \det M = \det M^\dag,\quad i\in 1,2 .
\end{equation}
This relation is satisfied by isospin chemical potential term, $\ol \psi \tau^3 \gamma^0 \psi$, ensuring that the determinant of the action is real.  Note, that the baryon chemical potential coupled to $\ol \psi \gamma^0 \psi$ will not satisfy this relation.  

Second,  in a finite volume, spontaneous symmetry breaking without an explicit symmetry breaking source cannot occur, unless there are infinitely many degrees of freedom.  The way to explore spontaneous symmetry breaking phases on the lattice is to include a small source of explicit symmetry breaking and extrapolate to when the source is zero.  This is the case when exploring the chiral condensate since simulations often have an explicit chiral breaking mass term along with any chiral breaking discretization effects.  While the source for chiral symmetry breaking is inherent to the system, the same cannot be said for a pseudoscalar condensate.  Thus, in order to measure the condensate, a small breaking term is required to ``tip" the condensate in a specific direction the the $\tau^1$-$\tau^2$ plane.  The pseudoscalar source term added to the Lagrangian is of the form \cite{Bedaque:2009yh} 
\begin{equation}\label{eq:L_Tip}
\D\cL = i \epsilon \ol \psi_x \gamma^5\frac{\tau^2}{2} \psi_x,
\end{equation}
which resembles a twisted mass term with a $\tau^2$ flavor coupling.  In addition to dictating the direction of the pseudoscalar condensate, this term also is necessary to ensure the Dirac inversion does not become singular \cite{Kogut:2002zg,Kogut:2004zg}.  If one accounts for this term in the \CPT Lagrangian, the minimum of the potential in Eq.~\eqref{eq:L_Cont_Pot} \cite{Son:2000xc,Son:2000by}
\begin{equation}
\label{eq:Pot_Min}
\cos \alpha = \frac{2Bm_q}{\mu_I^2}-\frac{B\epsilon}{\mu_I^2}\cot \a.
\end{equation}
As long as $\e$ is small (taken to zero before $m_q$ or $\mu$), the continuum, infinite volume result is recovered.  However, it is worth noting that if $\e$ is not sufficiently small, the $\cot \a$ term diverges as the critical limit is approached from above, rendering the expansion invalid.   The resulting vacuum expectation value (talking $\e \rightarrow 0$ first) leads to an altered chiral condensate and a pseudoscalar condensate given by
\begin{eqnarray}\label{eq:cond}
\langle \ol \psi \psi \rangle &=&F^2 B \cos \a, \notag \\
i \langle \ol \psi \frac{\tau^2}{2} \gamma_5 \psi \rangle &=& F^2 B \sin \a.
\end{eqnarray}

Third, while discretization effects are less pronounced for the isospin chemical potential case than chiral symmetry breaking, care needs to be taken when performing tunings or interpreting residual mass terms (especially when taking $m_q$ to zero).  The system now has a pseudoscalar condensate and a chemical potential which lead to several of the alterations of the usual tuning methods (such as ensuring the pion mass vanish in the chiral limit) as will be described in Sec.~\ref{sec:Confine}.  Additionally, the isospin chemical potential term should correspond to the time component of the conserved vector current.  Thus, in certain formulations, such as Domain Wall fermions, the used of the local vector current may not yield the desired results.  However, this concern appears to be insignificant in lattice studies of small baryon chemical potentials \cite{Hegde:2008nx}.

\section{Confining Theories}     \label{sec:Confine}       
Confining theories in the context of this paper are theories where dynamical symmetry breaking occurs and a confinement scale emerges.  One class of theories similar to QCD, has scales that follow the hierarchy (including the isospin chemical potential in the range of interest)
\begin{equation}
\label{eq:Confine_Hier}
\frac{1}{L} \lesssim m_q \lesssim \mu_I \ll \Lambda_{QCD} \lesssim \frac{1}{a}, 
\end{equation}
where $L$ is the spacial extent of the lattice and $a$ is the lattice spacing.  In this particular scenario, the scale $\L_{QCD}$ sets both the IR confinement scale that observables, such as nucleons and rho mass, are proportional to and the UV scale when the asymptotically free theory becomes perturbative. As a result, these theories are thought to be phenomenologically similar to QCD.

Another possible class of confining theories for a different set of parameter (often with a greater number of flavors), usually referred to as walking theories, follow a different hierarchy
\begin{equation}
\label{eq:Walk_Hier}
\frac{1}{L} \lesssim m_q \lesssim \mu_I \ll \Lambda_{IR} \ll \Lambda_{UV} \lesssim \frac{1}{a}. 
\end{equation}
Most notably, the primary difference is the separation of the IR confinement scale and the UV perturbative scale.  For a given renormalization scheme, the running of the coupling slows between the two scales, and this slow running is believed to give  an enhancement to the chiral condensate proportional to $(\Lambda_{UV}/\Lambda_{IR})^\nu$, where $\nu$ is a power determined by the dynamics and renormalization scheme \cite{Appelquist:1986an}.  For this reason, extracting the chiral condensate in the chiral limit (or, in terms of the \CPT parameters, $B/F$) is the main priority for theories in the confining regime.  

For both classes of confining theories, \CPT extrapolations are valid provided both $m_q$ and $\mu_I$ are small enough.  However, how small they have to be for a valid chiral extrapolation is expected to be significantly different between the two theories.  There are two primary reasons for this.  First, walking theories usually require a large number of flavors. Powers of the factor $N_f$ appear in the coefficients  of the higher order chiral logarithms and these coefficients can lead the higher order terms to be competitive with lower order terms if the pion mass is not small enough. This would result in a breakdown of the chiral EFT.   Second, the pion mass itself is enhanced in these theories, so an even smaller quark mass is required to make a pion mass in the chiral regime.  As a result, multiple lattice calculations of ``confining" theories have found that chiral extrapolations do not hold for their current quark masses \cite{Neil:2010sc} (if running slow enough, such theories may resemble an IR conformal theory where anomalous dimension is fixed, $\gamma(\mu) \approx \gamma^*$, and the scaling analyses in Sec.~\ref{sec:Conform} may prove informative). It should be emphasized that a small isospin chemical potential (compared to the chiral breaking scale) is required for the following relations to hold.

The study of the effects of isospin chemical potentials for QCD has been studied extensively for masses and condensates \cite{Son:2000by,Kogut:2001id,VillavicencioReyes:2004pq} and the Dirac spectrum \cite{Kanazawa:2011tt}.  This section will reflect a small subset of these effects as it pertains to extracting useful information for QCD-like theories.  
\subsection{Pion mass}                %

The effect of the isospin chemical potential on the pion mass is separated in two regions, when the isospin chemcial potential is too small to form a pseudoscalar condensate ($\mu_I^2 < 2 B m_q$) and when it is large enough that a pseudoscalar condensate is formed ($\mu_I ^2> 2Bm_q$).  Both cases have been studied extensively \cite{Son:2000by,Kogut:2001id,VillavicencioReyes:2004pq} and the relevant tree-level results for extracting information about the condensate are presented here.  Again, it should be noted that $2 Bm_q, \mu_I^2 \ll \Lambda_\chi^2$ is required for the following analysis to be valid.

In the first phase, $\mu_I^2 < 2 B m_q$, no pseudoscalar condensate is formed and the chiral Lagrangian follows the form of Eq.~\eqref{eq:L_Cont_CPT}.  Here, much like zero density, the propagator matrix is has no cross terms (only terms proportional to $\pi_0^2$ or $\pi_+\pi_-$) and the particle masses are given by the zero values of its determinant
\begin{equation}
\det \mathbf{D}^{-1} = [p^2 -  2 B m_q][(p-\mu_I)^2- 2 B m_q][(p+\mu_I)^2- 2 B m_q] = 0,
\end{equation}
and the resulting pion masses are  (assuming the sign of $\mu_I$ is defined by $\mu_I = -|\mu_I|$)
\begin{equation}\label{eq:mass_no_cond}
m_{\pi_+} = \sqrt{2B m_q} - |\mu_I| \quad,\quad  m_{\pi_0} = \sqrt{2B m_q} \quad,\quad m_{\pi_-} = \sqrt{2B m_q} + |\mu_I|. 
\end{equation}
In effect, at zero temperature, this phase only results in a shift in energy of the charged pions \cite{Cohen:2003kd}.  While information about the condensate can be extracted from this phase, it is the same information that can be extracted from zero density simulations.

\begin{figure}[t]
   \centering
   \includegraphics[width=120mm]{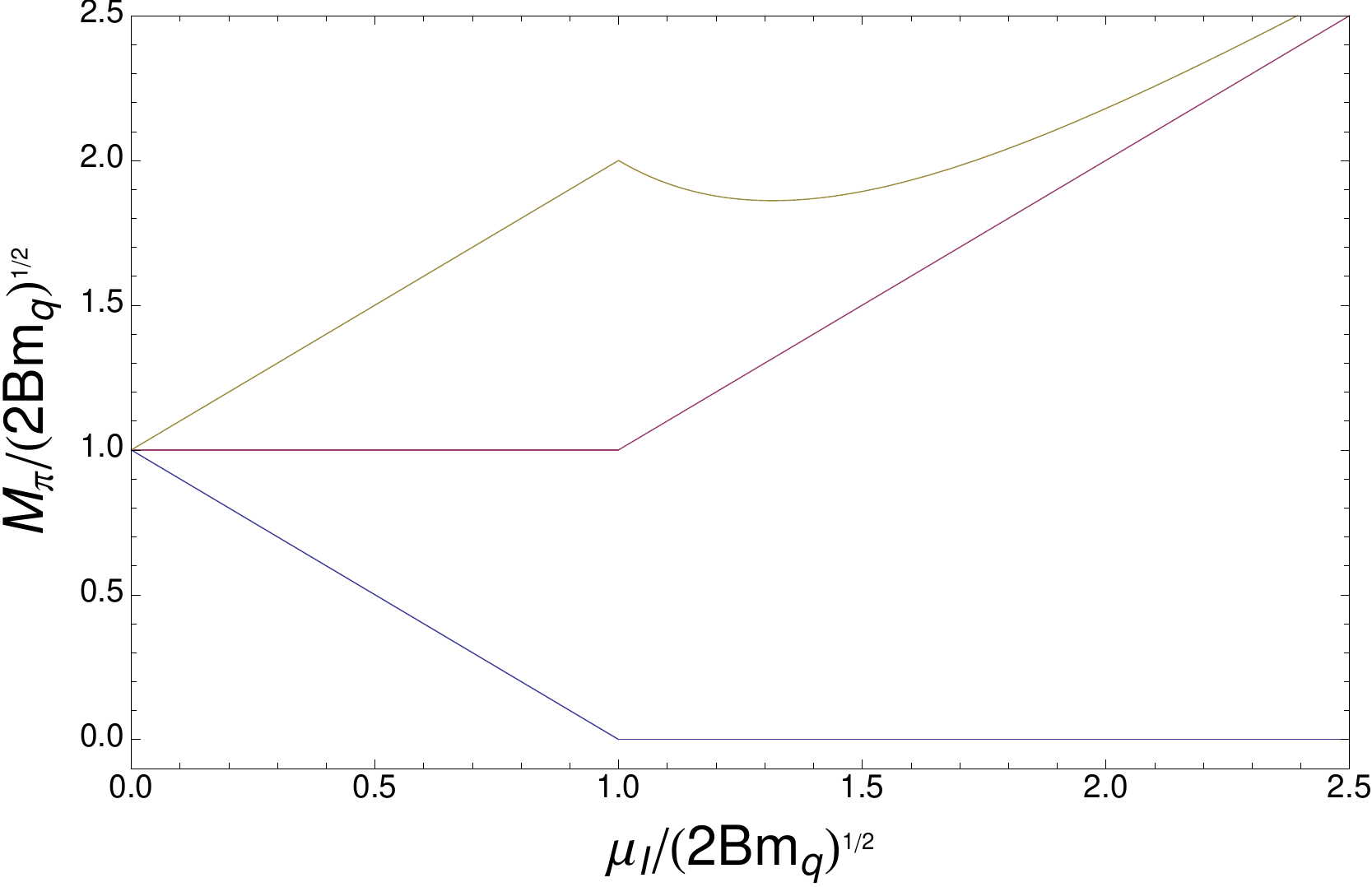} 
   \caption{Pion masses measured  vs. (negative) isospin chemical potential.  All axis are normalized to the LO vacuum pion mass value $\sqrt{2 B m_q}$.   The top curve ($m_{\pi_-} $), middle curve  ($m_{\pi_0} $), and bottom curve ($m_{\pi_+}$) are defined in Eq.~\eqref{eq:mass_no_cond} and Eq.~\eqref{eq:mass_cond}.  The curvature of $m_{\pi_-} $ after the pseudoscalar condensate forms ($\mu_I^2 > 2Bm_q$) can give additional information useful for extracting the chiral condensate.}
   \label{fig:pion_masses}
\end{figure}

In the second phase,  $\mu_I^2 > 2 B m_q$, a pseudoscalar condensate is formed and the chiral Lagrangian is now expanded about a new vacuum under the transformation $\Sigma \rightarrow \xi_0 \tilde{\Sigma} \xi_0$ where $\Sigma_0 =  \xi_0^2 = \cos \a + i \mathbf{n}\cdot\bm{\tau}\sin\a$.  This leads to a propagator matrix that has no cross terms for the neutral particle and cross terms (proportional to $\pi_+ \pi_+$ or $\pi_- \pi_-$) for the charged particles due to the flavor mixing  of the pseudoscalar condensate.  The zeros of the determinant (with three-momenta set to zero) are given by   
\begin{equation}
\det \mathbf{D}^{-1} = [p_0^2 -  \mu_I^2][p_0^2- \mu_I^2(1+3(2 B m_q/\mu_I^2)^2]p_0^2 = 0,
\end{equation}
and the resulting pion masses are given by 
\begin{equation}\label{eq:mass_cond}
m_{\pi_+} = 0 \quad,\quad  m_{\pi_0} =  |\mu_I| \quad,\quad m_{\pi_-} =  |\mu_I| \sqrt{1+3\bigg(\frac{2Bm_q}{\mu_I^2}\bigg)^2}.
\end{equation}
The first two masses do not contain any information about the chiral condensate, but the mass of the third pion is dynamically altered (as compared to the zero density case) as a function of the condensate and the isospin chemical potential.  In Fig.~\ref{fig:pion_masses}, this can be interpreted as the curvature in $m_{\pi_-}$ when $\mu_I^2 > 2 B m_q$. The usual Gell-Mann-Oakes-Renner (GOR) relation ($m_\pi^2 f_\pi^2 = 2 m_q \langle \ol \psi \psi \rangle$) no longer holds in the presence of the pion condensate, and the mass of $m_{\pi_-}$ leads to a modified GOR-like relation given by
\begin{equation}
\frac{B}{F} = \sqrt{\frac{\mu_I^4}{12m_q^2F^2}\bigg(\frac{m_{\pi_-}^2}{\mu_I^2}-1\bigg).} 
\end{equation}
The primary advantage gained from this calculation (as pointed out in Ref.\cite{Bedaque:2009yh}) is the addition of another ``knob" besides $m_q$ to extract chiral behavior.  Thus, for each $m_q$ within the chiral regime, multiple $\mu_I$ also within the chiral regime can be calculated (whose costs are determined by the symmetry breaking term $\epsilon$), and this information can lead to more control of chiral extrapolations.
\subsection{Pseudoscalar condensate }                %
In addition to extracting $B/F$ from the pion mass, another approach would be to focus on measuring the pseudoscalar density directly.  This has been performed stochastically  in both quenched \cite{Kogut:2002tm} and dynamical \cite{Kogut:2002zg,Kogut:2004zg,deForcrand:2007uz} QCD at finite isospin density.  In stochastic calculations of the chiral condensate, one nuisance is the presence of a large divergent piece on the order of the cut off scale, which masks the finite physics of interest.  This can be seen in the naively discretized free propagator loop
\begin{equation}
\label{eq:Naive_divergence}
\tr \int_{-\frac{\pi}{a}}^{\frac{\pi}{a}} \frac{d^4p}{(2\pi)^4}\frac{1}{\frac{i}{a}\gamma_\mu\sin a p_\mu - m_q} = 2m_q\int_{-\frac{\pi}{a}}^{\frac{\pi}{a}}\frac{d^4p}{(2\pi)^4} \frac{1}{\frac{1}{a^2}\sin^2 a p_\mu - m_q^2} \rightarrow \frac{2m_q}{a^2},
\end{equation}
where the momentum dependence in the numerator is zero due to the trace over the $\gamma$-matrix. For the pseudoscalar condensate, a similar lattice artifact given by
\begin{equation}
\label{eq:Naive_divergence}
\tr \int_{-\frac{\pi}{a}}^{\frac{\pi}{a}} \frac{d^4p}{(2\pi)^4}\frac{\gamma^5 \tau^2}{\frac{i}{a}\gamma_\mu\sin a p_\mu + m_q+i\epsilon\gamma^5\tau^2} = 2\epsilon\int_{-\frac{\pi}{a}}^{\frac{\pi}{a}}\frac{d^4p}{(2\pi)^4} \frac{1}{\frac{1}{a^2}\sin^2 a p_\mu + m_q^2+\epsilon^2} \rightarrow \frac{2\epsilon}{a^2}.
\end{equation}
This divergent contribution will be smaller than its chiral condensate counterpart as long as $\epsilon < m_q$.  Neither divergence should depend on $\mu_I$ and consequently, taking differences at different $\mu_I$ may prove valuable. 

\begin{figure}[t]
   \centering
   \includegraphics[width=120mm]{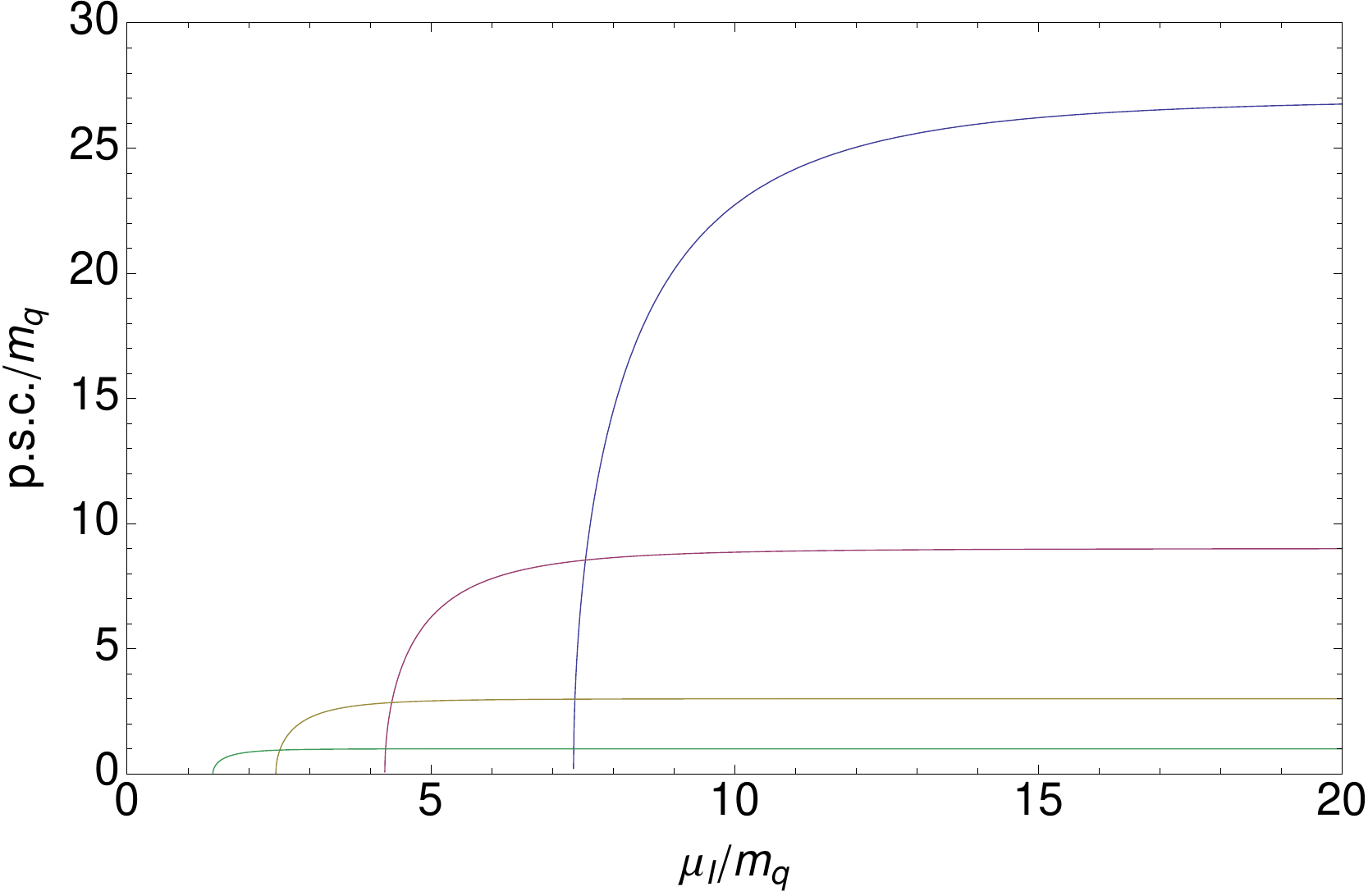} 
   \caption{Pseudoscalar condensate (p.s.c) vs. isospin chemical potential (both normalized to a fixed small $m_q$) for theories with $B/m_q$ values of  1 (green), 3 (orange), 9 (magenta) and 27 (blue).  As the chiral condensate increases, higher $\mu_I$ values are required to form a pseudoscalar condensate.}
   \label{fig:Pseu_conf}
\end{figure}

Following up Eq.~\eqref{eq:cond}, the pseudoscalar condensate at LO in the chiral expansion ($m,\mu_I \ll \L_\chi$) is given by
\begin{equation}
\label{eq:Pseudo_cond}
i \langle \ol \psi \frac{\tau^2}{2} \gamma_5 \psi \rangle = F^2 B \sin \a = F^2 B\sqrt{1-\Bigg(\frac{2Bm_q}{\mu_I^2}\Bigg)^2}.
\end{equation}
The resulting pseudoscalar condensate as a function of isospin chemical potential for fixed $m_q$ and various values of $B$ is shown in Fig.~\ref{fig:Pseu_conf}.  Ultimately, we will want to take $m_q$ to zero (more specifically, take $2Bm_q/\mu_I^2$ to zero).  Upon doing this, the pseudoscalar condensate reduces to the quantities of interest, namely
\begin{equation}
\label{eq:Pseudo_cond_lim}
\lim_{m_q\rightarrow 0} i \langle \ol \psi \frac{\tau^2}{2} \gamma_5 \psi \rangle = F^2 B. 
\end{equation}
It should be noted that this particular result does rely on the fact that $\mu_I \ll \L_\chi$, and corrections on the order of $\mu_I^2$ are to be expected.  However, as long as $\mu_I$ is well within the chiral regime, this particular method gives a clear method of extracting $B$ from multiflavor lattice calculations with the large lattice artifact contributions potentially reduced.  Also, getting at the chiral value for the pseudoscalar decay constant can be just as hard (if not harder) than extracting the chiral condensate.  Thus, by combining this analysis with the pion mass analysis from the previous sections, $B$ and $F$ can be determined separately.

\section{IR Conformal Theories} \label{sec:Conform}         
Upon reaching a certain critical value for the number of flavors in a theory with a given number of colors and fermions in a given representations, the theory is believed to be IR conformal due to the presence of an IR fixed point.  While the theory still remains asymptotically free for large energies, the fixed value for the coupling prevents any IR confinement scale from emerging.  As a result, the theory does not have any spontaneous symmetry breaking, and in the absence of any explicit massive scale, all observables have zero mass.  For this reason, low-energy effective field theories that depend on spontaneous chiral symmetry breaking, such as \CPT, are no longer valid.  In the case of a mass deformed IR conformal theory, where a small quark mass is added ($m_q \ll \L_{UV}$), all hadronic masses follow the relation \cite{DeGrand:2009mt,DelDebbio:2010jy,DelDebbio:2010ze,Appelquist:2009ty}
\begin{equation}
\label{eq:Mass_Conform }
M_\cO \sim m^{\frac{1}{1+\gamma^*}},
\end{equation}
where $\gamma^*$ is the scheme-independent mass anomalous dimension.  Many phenomenological arguments for understanding the quark flavor hierarchy from technicolor models that are ``nearly" IR conformal ($\gamma(\mu) \approx \gamma^*$) are based on the magnitude of $\gamma^*$ \cite{Chivukula:2010tn} and for this reason, it is the primary quantity of interest for IR conformal theories.  However, many intricacies occur when extending this to the lattice.  The first is that the relation \label{eq:Mass_Conform} only holds for small quark masses such that higher order terms are negligibly small.  Another related concern is that in addition to the mass deformation, the inverse of the finite spacial extent, $L^{-1}$, acts as an IR cutoff to the theory and the hierarchy $L^{-1} \ll a^{-1}$ should be maintained to minimize the contributions from unphysical UV artifacts.

Upon including the isospin chemical potential, the relevant hierarchy of scales in the conformal window are given by
\begin{equation}
\label{eq:Conform_Hier}
\frac{1}{L} \lesssim m_q \lesssim \mu_I  \ll \Lambda_{UV} \lesssim \frac{1}{a}. 
\end{equation}
As will be shown in the subsequent sections, the implementation of the isospin chemical potential does not receive any renormalization due to the fact that it is coupled to a conserved current (the operator it is coupled to has engineering dimension as shown in Ref.~\cite{DelDebbio:2010jy}).  This feature allows for both varying the fermion mass near zero and allows for one to define finite density scaling relations akin to the finite size scaling.
 
\subsection{Universal scaling function}                %
The use of finite size scaling relations remains the best method for defining the universal scaling behavior and extracting the critical exponents for theories with an IR fixed point.  This fact has not escaped the lattice community and the scaling behavior has been worked out for multiple classes of observables and operators \cite{DeGrand:2009mt,DelDebbio:2010jy,DelDebbio:2010ze}.  While finite size scaling has been explored on the lattice \cite{DeGrand:2009et,DelDebbio:2010jy,DeGrand:2009hu,DeGrand:2011cu}, it ultimately proves to be quite demanding numerically since it requires multiple ensembles of different parameters while keeping $m_q$ small and $L^{-1} \ll a^{-1}$.   

A similar set of universal scaling curves can be derived for the isospin chemical potential since the isospin chemical potential does not change under renormalization group (RG) transformations as a result of chemical potentials being coupled to conserved currents in the Lagrangian.  The analysis for this ``finite density" scaling (assuming the volume is large and fixed) is essentially the same as the analysis performed for finite size scaling in Ref.~\cite{DelDebbio:2010jy}, with the key difference being the spacial extent $L$ is replaced with the isospin chemical potential $\mu_I^{-1}$.  Nevertheless, the key elements to this analysis will be repeated here.

The correlation function calculated on the lattice is given by
\begin{equation}
\label{eq:Corr}
C_\cO(t;m_q,\mu_I,L,\mu) = \langle \cO^\dag(t) \cO(0) \rangle \rightarrow A e^{-M_\cO t},
\end{equation}
where $\cO$ is an artibrary interpolating operator and $\mu$ is the renormalization scale (not to be confused with the isospin chemical potential, $\mu_I$) and the r.h.s represents the long Euclidean time behavior of the correlation function.  Assuming a fixed value for $L$ with $L^{-1} \ll m_q, \mu_I$ and defining $\hat{m}_q = m_q/\mu$, one arrives at the correlation function $C_\cO(t;\hat{m}_q,\mu_I,\mu)$.  Under the RG transformation $\mu \rightarrow \mu^\prime$, where $\mu = b \mu^\prime$, the correlation function becomes
\begin{equation}
\label{eq:Corr_RG}
C_\cO(t;\hat{m}_q,\mu_I,\mu) = b^{-2\gamma_\cO} C_\cO(t;b^{1+\gamma^*}\hat{m}_q,\mu_I,\mu^\prime).
\end{equation}
where $\gamma_\cO$ is the anomalous dimension of the field operator $\cO$.  Performing a na\"ive dimensional scaling by $b$ to this result, 
\begin{equation}
\label{eq:Corr_Naive}
C_\cO(t;\hat{m}_q,\mu_I,\mu) = b^{-2(\gamma_\cO+d_\cO)} C_\cO(b^{-1} t;b^{1+\gamma^*}\hat{m}_q,b \mu_I,\mu).
\end{equation}
Now choosing $b$ such that $b \mu_I = U_0$ where $U_0$ is a dimensionless constant.   This leads to the correlation function
\begin{equation}
\label{eq:Corr_Def}
C_\cO(t;\hat{m}_q,\mu_I,\mu) = \Bigg(\frac{U_0}{\mu_I}\Bigg)^{-2(\gamma_\cO+d_\cO)} C_\cO\Bigg( \frac{\mu_I}{U_0} t;x \frac{U_0^{1+\gamma^*}}{\mu}, U_0,\mu\Bigg),
\end{equation}
where $x = m_q/\mu_I^{1+\gamma^*}$ and from the asymptotic behavior of the correlation function at long Euclidean time, the corresponding mass is given by a function of $x$,
\begin{equation}
\label{eq:Mass_Scale}
M_\cO \sim \mu_I F(x).
\end{equation}
In the same way varying the box size is used in finite size scaling techniques, the relevant information of the mass anomalous dimension can be extracted by varying $\mu_I$ as the fermion mass approaches zero.  In particular, when plotting $M_\mathcal{O}/\mu_I \ \text{vs.} \ m_q/\mu_I^{1+\gamma^*}$ at different $\mu_I$ values, the anomalous dimension can be determined by varying $\gamma^*$ and observing when all data points fall on the same curve.  While this is akin to finite size scaling, the isospin chemical potential has more resolution at a comparable cost (can be adjusted ala quark mass without critical slow down) than the volume, which usually requires additional lattice sites (expensive, discrete) or tuning the lattice spacing (additional lattice artifacts). 

Using similar techniques, volume scaling can be incorporated into Eq.~\eqref{eq:Mass_Scale}.  Following the same procedure as above, any mass will now be a universal function of two variables, $M_\cO \sim \mu_I F(x,y)$, where $x = m_q/\mu_I^{1+\gamma^*}$ and $y = \mu_I L$.  
\subsection{Condensates}                %
As in confining scenarios, condensates can give important information regarding IR conformal theories.  In particular, the study of condensates in the presence of a chemical potential in a given theory can provide another test as to whether or not a theory is conformal or confining.  In an IR conformal theory with $\mu_I=0$ and $m_q = 0$, all observables with mass dimension, including chiral and pseudoscalar condensates will be zero (technically, all observables should scale with $L^{-1}$ since the finite spacial extent is a relevant operator in an IR conformal theory).  The interesting question is the behavior of these observables at non-zero (but small) values of $\mu_I$ and $m_q$, and, in particular, how they compare to their confining counterparts.

While spontaneous symmetry breaking does not occur in IR conformal theories, spontaneous symmetry breaking also does not occur without a small source for confining theories at finite volume.  Thus, as explained in Sec.~\ref{sec:Iso}, a small explicit symmetry breaking parameter (which we called $\e$) must be introduced to the system for the pseudoscalar condensate at $\mu_I > 2Bm_q$ to be measured in both confining and IR conformal theories.  However, the phenomena between these two theories should behave differently when the limit $\e \rightarrow 0$ is taken first.  From Eq.~\eqref{eq:Pseudo_cond}, 
it is evident that as $m_q \rightarrow 0$, the pseudoscalar condensate is proportional to the confinement scale (in particular, the chiral condense).  This  is to be expected, since spontaneous flavor symmetry breaking is expected to occur in infinite volume for theories with a large enough isospin chemical potential.  However, the more interesting aspect stems from the fact that as $m_q \rightarrow 0$, the pseudoscalar condensate has a diminishing dependence on $\mu_I$.


Within the conformal window, the pseudoscalar condensate takes a significantly different form when $m_q$ and $\mu_I$ are small, but non-zero, where $m_q^{1/(1+\gamma^*)}$ plays the role of the confinement scale.  Comparing the condensates in the chiral limit, 
\begin{equation} \label{eq:scal_fun}
 \lim_{m_q\rightarrow 0} i \langle \ol \psi \frac{\tau^2}{2} \gamma_5 \psi \rangle\rightarrow
\begin{cases}
 f^2 B, & \text{Confining Theory}, \\
D \,\mu_I^3, &\text{IR Conformal Theory},
\end{cases}
.\end{equation} 
where $D$ is a dimensionless constant that could be zero.  While this quantity is not an order parameter (unless $D=0$), there is  a difference in the behavior of the condensate as the isospin chemical potential is varied.  By reducing the fermion mass to small values, the analysis of the pseudoscalar  condensate is expected to display  different behavior at several different (but small) isospin chemical potential values.  Once again, it should be made clear how much this process depends on orders of limits.  First, the explicit pseudoscalar breaking parameter $\e$ needs to be taken to zero, followed by $m_q$, while keeping $\mu_I$ small enough such that these results will not be contaminated from higher order effects.

\section{Estimates of lattice artifacts}  \label{sec:Lat_Art}          
In addition to the increased computational cost of inversions from reducing the quark mass, one always needs to be aware of the lattice artifacts that can arise.  For example, simply decreasing the quark mass  leads to $m_\pi L$ decreasing as well, which can introduce significant lattice artifacts, especially if  $m_\pi L < 4$.  It should be noted that while this is a limitation for confining theories, large volume effects can act as a feature for finite-size scaling in IR conformal theories (more defined volume behavior). In addition to large IR effects, significant UV effects can become important, such as the emergence of non-physical phases such as the Aoki phase \cite{Aoki:1983qi,Sharpe:1998xm}.  In this section we will address both of these examples when the quark mass is dropped in the presence of a small isospin chemical potential.  The following two sections will address these lattice artifacts for the confining scenario only.

\subsection{Finite Volume}                %
The dominant volume effects in a lattice system with periodic boundary conditions are set by the mass of the lightest particle mass times the finite spatial extent of the system (for most theories, this is set by the product $m_\pi L$).   In a confining QCD-like system, for small enough quark masses, \CPT can be used to estimate the volume effects due to the finite spacial extent \cite{Colangelo:2003hf}.  These effects emerge in the calculation of quantum corrections and loop diagrams, which the continuous momenta integrals are replaced by sums.  In particular, for a periodic box,
\begin{equation} \label{eq:box_sums}
\int \frac{d^3 p}{(2\pi)^3} \rightarrow \frac{1}{L^3}\sum_{\frac{2\pi\mathbf{n}}{L}}
.\end{equation} 
The resulting finite volume effects for the pion propagator are often exponentially suppresses, as in the one loop correction term that appears in the calculation of the pion mass and chiral condensate  \cite{Gasser:1983yg,Gasser:1984gg,Colangelo:2003hf,Bedaque:2006yi}
\begin{eqnarray} \label{eq:finite_one_loop}
\int \frac{dp_0}{2\pi}\Bigg[\frac{1}{L^3}\sum_\mathbf{n}-\int \frac{d^3p}{(2\pi)^3} \Bigg]  \frac{i}{p^2 - m_\pi^2} &=& \frac{m_\pi}{4\pi^2L}\sum_{\mathbf{n} \neq 0} \frac{1}{|\mathbf{n}|}K_1(|\mathbf{n}|m_\pi L)\nonumber\\
 &\rightarrow&  \frac{m_\pi}{4\pi^2L}\sum_{\mathbf{n} \neq 0} \frac{1}{|\mathbf{n}|^{3/2}}e^{-|\mathbf{n}| m_\pi L}
,\end{eqnarray} 
where the asymptotic form of the Bessel function $K_1(x) \simeq e^{-x}/\sqrt{x}$.  As the number of flavors is increased, the coefficient of this exponential is also enhanced.  In the presence of an isospin chemical potential, the volume effects are subtle.  In particular, when $\mu_I \gg m_q$, two of the modes have $m_\pi \sim \mu_I$, leading to heavier particles.  However, the more noteworthy case is that in the condensed phase, one of the pseudoscalar particles becomes a massless Goldstone mode, whose correlation length is infinite and will always have volume effects.  In practice, the mass of this Goldstone mode is cut off by the explicit flavor breaking term $\epsilon$ that is defined in Eq.~\eqref{eq:L_Tip} leading to $m_\pi \sim \epsilon$.  The corresponding volume effects depend upon the quantity $\epsilon L$, where there are two separate limits.  The limit $\epsilon L \rightarrow \infty$ represents taking the box to infinity before taking the explicit flavor breaking to zero.  This limit is likely too computationally expensive to ensure.  The second limit $\epsilon L \rightarrow 0$ while all other mass scales times the box goes to infinity will always be sensitive to the size of the system.  However, if the box size is large enough, the system should be in the ``epsilon regime" where the zero-mode volume effects are enhanced. Thus, epsilon regime chiral perturbation theory or random matrix theory can accurately account for these volume effects \cite{Gasser:1987ah,Hansen:1990yg,Bedaque:2004dt,Smigielski:2007pe}.  Additionally, mesonic two-point functions in the presence of an isospin chemical potential were previously calculated in the epsilon expansion \cite{Akemann:2008vp}.


\subsection{Lattice spacing and Aoki regime}                %
The other well known lattice phenomena that occurs when reducing the mass while keeping all other parameters fixed is the emergence of the Aoki regime  \cite{Aoki:1983qi,Sharpe:1998xm}, where flavor and parity are spontaneously broken.    This unphysical phase has been shown to depend on regions of the $m_0$-$g^2$ phase diagram and can be explained via a \CPT argument based on estimates of the lattice spacing effects \cite{Sharpe:1998xm}.  This phase can play a significant role when the LO \CPT effect (given by the leading order pion mass) becomes on the same order as the NLO $\cO(a^2)$ effect assuming the $\cO(a^2)$ effect comes in with the opposite sign.  These competing effects, as worked out in Ref.~\cite{Sharpe:1998xm}, are most apparent when minimizing the potential of the \CPT Lagrangian with the parameterization, $\S = A + i \mathbf{B} \cdot \bm{\sigma}$ 
\begin{equation} \label{eq:Aoki_Pot}
V(A) = -c_1 A + c_2 A^2
.\end{equation} 
where $c_1 \sim m_\pi^2$ and $c_2 \sim a^2$.  In most lattice calculations with unphysical large pion masses and reasonably small lattice spacings, $c_1 \gg c_2$ and these two effects never compete.  However, as the pion mass is dropped while the lattice spacings are fixed, scenarios when $c_1 \sim c_2$ can occur and new unphysical minima can emerge in the potential.  However, just because $c_1 \sim c_2$ does not mean that one is guaranteed to be in the Aoki regime when the actual lattice simulations are performed.

When an isospin chemical potential is included, as $m_q$ is dropped, $m_\pi \sim \mu_I$ or $m_\pi \sim \epsilon$.  Before, we needed to ensure that our lattice spacing was chosen such that $m_\pi^2 \gg a^2 \Lambda^4$.  Now,  $\mu_I^2 \gg a^2 \Lambda^4$ and $\Lambda \epsilon \gg a^2 \Lambda^4$ is needed to ensure no Aoki phase emerges.  It might appear that only $\mu_I^2 \gg a^2 \Lambda^4$ is needed to have $c_1 \gg c_2$, and while this may end up being be the case, the determination of the sign and magnitude of $c_1$ and $c_2$ is a subtle non-perturbative question (i.e. large strange quark in lattice QCD does not necessarily prevent Aoki regime).  It should also be noted that the Aoki regime and the condensed finite isospin density phase are quite similar in appearance, as they both produce a flavor and parity violating pseudoscalar condensate.  Thus, lattice simulations in the Aoki regime may be incorrectly interpreted as an isospin condensed phase.

Another aspect involving lattice spacing effects is the fact that irrelevant operators can lead to new mass-like terms in the Lagrangian proportional to $aW$ where $W\sim \Lambda^3$.  Thus, since the LO combination in \CPT is given by $Bm_q + aW$, simply taking $m_q$ to zero is not enough to remove the ``lattice mass".  However, this lattice spacing effect should be reduced for improved actions, chiral lattice actions, or smaller lattice spacings.  Nevertheless, negative values of $m_q$ may need to be used to acquire the ``$m_q \rightarrow 0$" behavior discussed throughout this paper.

\section{Conclusion}            

In this work, we have presented a new scenario for extracting information from strongly coupled gauge theories by implementing an isospin chemical potential.  Due to its positive definite determinant, this isospin chemical potential can be included in lattice calculations and has multiple properties that can be used to help extract desired lattice quantities.    First, when the isospin chemical potential is larger than its critical value, a pion condensate is formed.  This pion condensate has multiple properties in the chiral regime that are beneficial for extracting the chiral condensate.  Second, the isospin chemical potential couples to a conserved current and consequently, does not receive renormalization.  To that end, a finite isospin density scaling analysis can be employed to extract the mass anomalous dimensions using similar techniques to finite size scaling.  This scaling relation should hold for IR conformal and could also be applicable to slow running theories near conformality where the anomalous dimension is roughly constant, $\gamma(\mu) \approx \gamma^*$.  Additionally, multiple ensembles with different chemical potentials can be simulated at comparable costs (in contrast to increasing volume), which gives improved resolution to both chiral or scaling analyses. These properties  make a convincing case that this scenario should be pursued on the lattice.

While the benefits of employing an isospin chemical potential are plentiful, there are still multiple caveats that must be understood.  The first, and most significant, caveat is the fact that much of the analysis in this work depends greatly on the fact that  isospin chemical potential is in the chiral regime ($\mu_I \ll \Lambda_{IR}, \Lambda_{UV}$).  On the other end, for confining theories, volume effects are now governed by the combination $\mu_I L$ and $\epsilon L$, and analytical techniques in the $\epsilon L \rightarrow 0$ limit might be needed.  Also, contributions from Aoki phases are also possible and should be taken into account.  

One could make the observation that the inclusion of an isospin chemical potential while taking $m_q$ and $\epsilon$ to zero is simply moving the previous issues with the quark mass to the chemical potential.  In many ways, this astute point is indeed correct and for that reason, this method should be viewed as another tool for probing lattice phenomena.  However, the two other key features of the isospin chemical potential, namely the formation of a pion condensate and coupling to a conserved current, make this system inherently different from the usual lattice system with just a quark mass.  The majority of the interesting physics for these systems not only occur in the chiral limit, but greatly depend on how the theory varies near the chiral limit.  With the presence of the isospin chemical potential, these systems can be explored in exactly this way.

\begin{acknowledgments}
The author would like to thank Paulo Bedaque, Aleksey Cherman, Michael Cheng, Tom Cohen, Tom DeGrand, Luigi Del Debbio, George Fleming, Masanori Hanada, Tom Luu, Agostino Petella, David Schaich, Brian Tiburzi, Pavlos Vranas, and Joe Wasem for helpful insight and discussions.  The author would also like to thank Simon Hands for informative comments and observations.
This work was performed under the auspices of the U.S. Department of Energy by LLNL under Contract No. DE-AC52-07NA27344.  This research was partially supported by the LLNL LDRD ÒUnlocking the Universe with High Performance ComputingÓ 10-ERD-033. 
\end{acknowledgments}

\bibliography{techni-iso-chem} %

\end{document}